\documentclass[lettersize,journal]{IEEEtran}
\usepackage{amsmath,amsfonts}
\usepackage[linesnumbered,ruled,vlined]{algorithm2e}
\usepackage{array}
\usepackage{textcomp}
\usepackage{stfloats}
\usepackage{url}
\usepackage{verbatim}
\usepackage{graphicx}
\usepackage[caption=false,font=footnotesize]{subfig} 
\usepackage{cite}

\begin{document}

\title{NeuroPilot: A Real-time Brain-Computer Interface system to enhance concentration of students in online learning}
\author{Asif Islam,  Farhan Ishtiaque, Md. Muhyminul Haque, Farhana Sarker, Ravi Vaidyanathan, and Khondaker Abdullah-Al-Mamun
}



\maketitle

\begin{abstract}
The prevalence of online learning poses a vital challenge in real-time monitoring of students’ concentration. Traditional methods such as questionnaire assessments require manual intervention, and webcam-based monitoring fails to provide accurate insights about learners' mental focus as it is deceived by mere screen fixation without cognitive engagement. Existing BCI-based approaches lack real-time validation and evaluation procedures. To address these limitations, a Brain-Computer Interface (BCI) system is developed using a non-invasive Electroencephalogram (EEG) headband, FocusCalm, to record brainwave activity under attentive and non-attentive states. 20 minutes of data were collected from each of 20 participants watching a pre-recorded educational video. The data validation employed a novel intra-video questionnaire assessment. Subsequently, collected signals were segmented (sliding window), filtered (Butterworth bandpass), and cleaned (removal of high-amplitude and EOG artifacts such as eye blinks). Time, frequency, wavelet, and statistical features were extracted, followed by recursive feature elimination (RFE) with support vector machines (SVMs) to classify attention and non-attention states. The leave-one-subject-out (LOSO) cross-validation accuracy was found to be 88.77\%. The system provides feedback alerts upon detection of a non-attention state and maintains focus profile logs. A pilot study was conducted to evaluate the effectiveness of real-time feedback. Five participants underwent a 10-minute session comprising a 5-minute baseline phase devoid of feedback, succeeded by a 5-minute feedback phase, during which alerts were activated if participants exhibited inattention for approximately 8 consecutive seconds. A paired t-test (t = 5.73, p = 0.007) indicated a statistically significant improvement in concentration during the feedback phase.

\end{abstract}


\section{Introduction}
\IEEEPARstart{C}{oncentration} monitoring has become essential to ensure learning effectiveness in online classes, whose adoption has been steadily increasing across different educational curricula \cite{adoptiononlinelearning1}. Prior research shows that learners are more prone to mind-wandering when they are watching lecture videos (being regarded as online learning scenario) than during live lectures (traditional offline classes) \cite{onlinevsofflinemindwander}. Thus, to track the engagement of students in educational and learning settings, conventional methods of developing questionnaires \cite{QuestionnaireCamelia2018} or camera-based monitoring \cite{WebcamPatil2024, WebcamHossen2023} are used. Although questionnaires are  useful for the objective evaluation of students, they only serve as a manual tool that requires human intervention and input. Conversely, webcam-based monitoring systems are limited to analyzing external behavioral cues such as facial expressions, gaze, or posture, and therefore lack the ability to provide direct neurophysiological insights into learners’ cognitive attention levels \cite{WebcamLimitation}, alongside inducing privacy issues \cite{WebcamlimitNair2024}. Additionally, vision-based models fail to preserve robustness in low-light conditions. For this reason, Brain-Computer Interface (BCI) has emerged as a powerful tool to perform cognitive load evaluation of students while watching lecture videos.

BCI is a form of neurotechnology that is leveraged to establish a direct connection between humans and machines. Specifically, electroencephalography (EEG) signals are increasingly being utilized for various use cases, including bionic intelligence\cite{BionicWei2025}, vehicle control \cite{VehicleBasedWang2023, GroundVehicleZhuang2021,Hatnet}, and other assistive technologies \cite{DirecSenseBCILin2021, FlashlightNetDang2024,HumanPhotographResponseKaongoen2020,HMSImprovingOnlineActive}. BCI has gained considerable recognition in the classification of attentive and non-attentive states \cite{MentalStatesLee2023,HMSAuditoryAttention,HMSCognitionLevel,HMSCognitionLevel2}, including distraction \cite{DistractionEducationChoi2025} and drowsiness\cite{DrowsinessReddy2022}. Despite the fact that this article is particularly oriented to enhance concentration levels, terms like focus, attention, concentration, and engagement, which  are different constructs overlapping each other, can be used interchangeably as the goal is to improve online learning for students using BCI. It has been seen from studies that the prefrontal cortex plays a crucial role in evaluating sustained attention \cite{frontalregion1Sahu2024,frontalbetaalsoKaushik2022,frontalZhang2021}. In the frontal region, the features contributing to concentration level detection are predominantly the changes in delta, theta, alpha, and beta band power. \cite{betabandpowerKo2017,frontalregion1Sahu2024,frontalbetaalsoKaushik2022}.
\IEEEpubidadjcol{} 

Current state-of-the-art works use EEG devices for laboratory experimentation that are not fully consumer-grade, which makes them not suitable for real-time implementation. Some of them are Emotiv Insight, Emotiv EPOC+, and OpenBCI Ultracortex. These devices lack dry electrodes, requires accurate placement across multiple scalp regions, not cost-effective and present setup challenges for consumers \cite{ReviewEEGdevices, ConsumergradeSabio2024}. Among consumer-grade devices, cost-effective options such as Muse, NeuroSky MindWave, and FocusCalm were studied. However, existing works provide limited details about their data collection strategies in online learning scenarios \cite{Neurosky1, Muse1, Muse2,Muse3}, leaving a gap in collecting ground truth (referring to correct labeling of neural data containing actual concentration and actual non-concentration states). Other consumer-grade devices from companies such as Cognionics and Wearable Sensing have been promoted for use in sports scenarios, but their high cost and large number of electrodes limit their suitability for consumer-friendly applications in this field. Vital gaps are seen in real-time implementation of BCI research articulated to track the concentration of students in learning environments. Huang et al. \cite{Real-timeAttentionHuang2025} focused on real-time cognitive monitoring to improve concentration time with neuro-feedback mechanism. However, the data collection strategy included mental arithmetic tasks, which is not an identical simulation of a learning environment, and the data validation was done using post regulation assessments such as the Sustained Attention to Response Task (SART) \cite{SARTRobertson1997}, which fails to validate the participants' engagement during the mental arithmetic tasks. Furthermore, the working algorithm of the single-channel EEG headband, along with the feature extraction and model training process, were not discussed, as their work focused more on evaluating the effects of neurofeedback rather than on developing the classification methodology. Conrad and Newman developed a strategy to use EEG signals for evaluating mind-wandering in online classes \cite{EEGMindWanderingConrad2021}. However, the use of only pre- and post-assessment questionnaires is insufficient to validate learners’ attention levels during the video. The work done by Rehman et al. \cite{StudentAttentionQLearningRehman2025} relied on publicly available datasets, which poses the same limitations as the earlier research gaps. The studies summarized in Table \ref{tab:summaryofpriorstudies} are selected to represent existing approaches for monitoring learner attention in online learning environments. Inclusion criteria are works that addressed learner engagement or attention in online or remote learning settings, and use of other related modalities (e.g., webcam-based, neurofeedback) whose limitations are being addressed in our proposed approach. Exclusion criteria included works focusing purely on offline classroom learning, cognitive tasks unrelated to learning (e.g., driving simulators), or studies that did not explicitly address attention/engagement measurement.
\begin{table}[h]
  \caption{Summary of Prior Studies}
  \centering
\renewcommand{\arraystretch}{1.2} 
  \begin{tabular}{|>{\centering\arraybackslash}m{1.6cm}|>{\centering\arraybackslash}m{2.9cm}|>{\centering\arraybackslash}m{3cm}|}
    \hline
    Work & Focus & Limitations \\
    \hline
    Camelia \cite{QuestionnaireCamelia2018} & Questionnaire to validate learner engagement & Requires manual interventions \\
    \hline
    Patil \cite{WebcamPatil2024} and Hossen \cite{WebcamHossen2023} & Computer Vision (webcam) to understand learner engagement  & Lack of insights in cognitive processing \\ \hline
    Huang \cite{Real-timeAttentionHuang2025} & Validating the effects of neurofeedback in increasing attention span & Lack of proper documentation on classification algorithm\\ \hline
    Conrad \cite{EEGMindWanderingConrad2021} & Evaluating mind-wandering of students in online classes using EEG 
    & Lack of proper data validation techniques\\ \hline
    Rehman \cite{StudentAttentionQLearningRehman2025} & Attention classification using their Deep Q-Learning model  & Computationally intensive and less feasible for real-time implementation \\ \hline
  \end{tabular}
  \label{tab:summaryofpriorstudies}
\end{table}

In this study, a eeg based attention classification framework is proposed to address the research gaps, with a focus on real-time implementation in online learning scenarios. A novel data acquisition and validation design through an intra-video questionnaire assessment was done to label the blocks as ground truths. The experimentation was done on classifying two cognitive states of students, namely attention and non-attention. Twenty participants were recruited to participate in data collection experiments containing two sessions, designated for the two classes of data and the device used is a FocusCalm headband. After that, several preprocessing techniques were employed to get clean data, followed by feature extraction, feature selection and classification. Following this, leave-one-subject-out validation was used in order to evaluate the classifier model when faced with unseen data. Finally, a simple app GUI was developed to have real-time insights of concentration levels along with a feedback mechanism to alert the learner during non-attention states. 
Thus, this article has contributions in the following areas:
\begin{itemize}
    \item A data collection paradigm is introduced that uses authentic lecture videos from educational technology platforms as EEG stimuli, offering a closer approximation to real-world online learning environments than traditional controlled tasks.
    \item A novel intra-video questionnaire assessment is proposed for data validation, which, unlike traditional post-task questionnaires or external measures (e.g., SART), collects learner responses during the lecture itself. This approach minimizes recall bias, provides time-specific ground-truth labels, and enhances the reliability of the EEG dataset.
    \item A real-time system is developed with feedback alerts to ensure the attention spans of learners improve over time. Preliminary evidence of its effectiveness in enhancing concentration is validated via a pilot study.
\end{itemize}

\section{Data Collection}
The section gives a detailed explanation of the inclusion criteria of participants, the device used for data collection, the experimental design, and setup.
\subsection{Participant Selection}
A total of 20 students with a mean age of $ 22.8 \pm 3.8 $ years were recruited for the data collection experiment. A detailed explanation of the experimental setup, collection paradigm, and the study was given to the participants before taking their informed consents. Participants retained the right to leave the data collection process at their discretion at any moment. This data collection protocol was approved by the Institutional Research Ethics Board (IREB), United International University (UIU). Among the participants, 15 are graduate students of Advanced Intelligent Multidisciplinary System Lab (AIMS Lab) under Institute of Research, Innovation, Incubation \& Commercialization (IRIIC), UIU and rest are the students of UIU. They were selected to pass the condition of having cognitive abilities necessary to comprehend a standard high school or equivalent curriculum-based educational material. This was ensured due to the video stimuli being exposed to the participants at the time of data collection, which contained online lectures on different topics prepared by national educational technology platforms. 
\begin{figure}[t]
    \centering
    \includegraphics[width=0.9\linewidth]{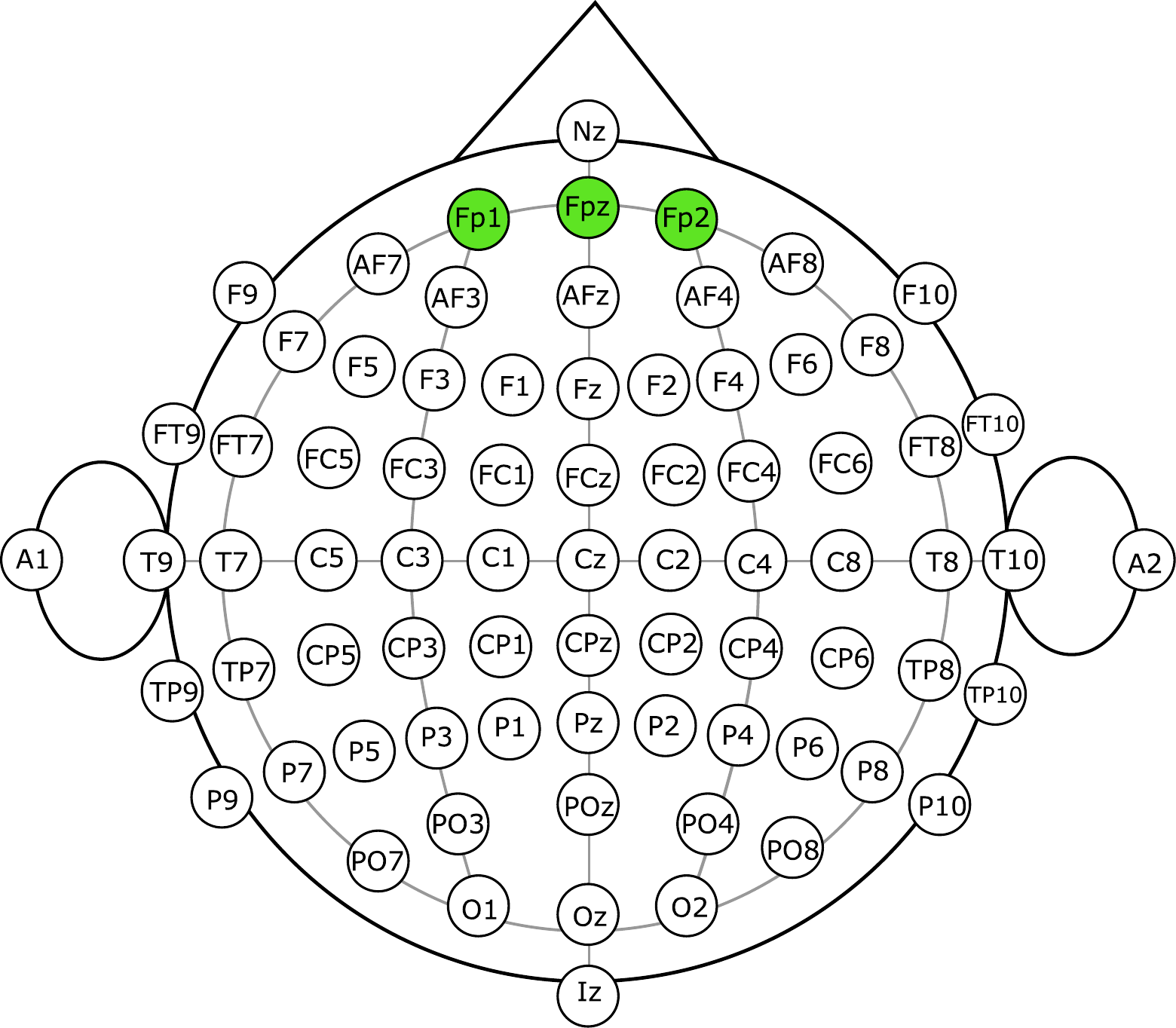}
    \caption{Positions of Fpz, Fp1 and Fp2 electrodes used by FocusCalm headband according to the international 10-10 electrode placement system.}
    \label{fig:placementelectrode}
\end{figure}
\subsection{Data Acquisition Device}
FocusCalm headband \cite{ConsumerGradeFocuscalmFlanagan2023} was used as the data acquisition device in this work for its consumer-grade accessibility, dry electrode system, frontal EEG coverage, low setup complexity and a balanced trade-off between signal quality and user comfort. It is a neuro-feedback EEG device designed for meditation purposes. There are three dry electrodes placed close to each other and aligns with the Fp1, Fp2, and Fpz positions of the international 10-10 electrode placement system as shown in Fig. \ref{fig:placementelectrode}. Fpz is used as the primary electrode, while Fp1 and Fp2 serve as the reference and ground electrodes \cite{focuscalmClinicsConsultants}. The device wirelessly sends single channel data at 250 Hz, along with eight distinct values, of which six indicate the delta, theta, alpha, low beta, high beta, and gamma frequency band values, and the other two show attention and meditation scores given by their developed model. The developers of FocusCalm at BrainCoTech have made a software development kit (SDK) named Crimson SDK, which is accessible for taking raw EEG data along with the eight other values in real-time.
\begin{figure*}[!t]
    \centering
    \subfloat[Data Collection Paradigm]{%
        \includegraphics[width=0.35\textwidth]{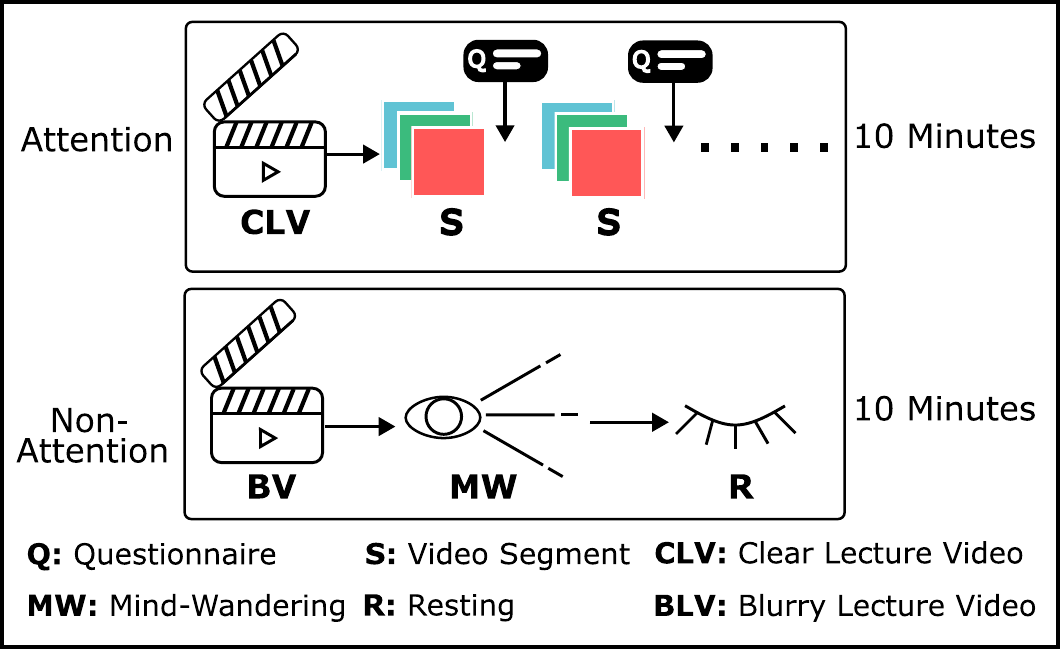}%
        \label{fig:datacollect}}
    \hspace{0.02\textwidth} 
    \subfloat[Data Validation Procedure]{%
        \includegraphics[width=0.49\textwidth]{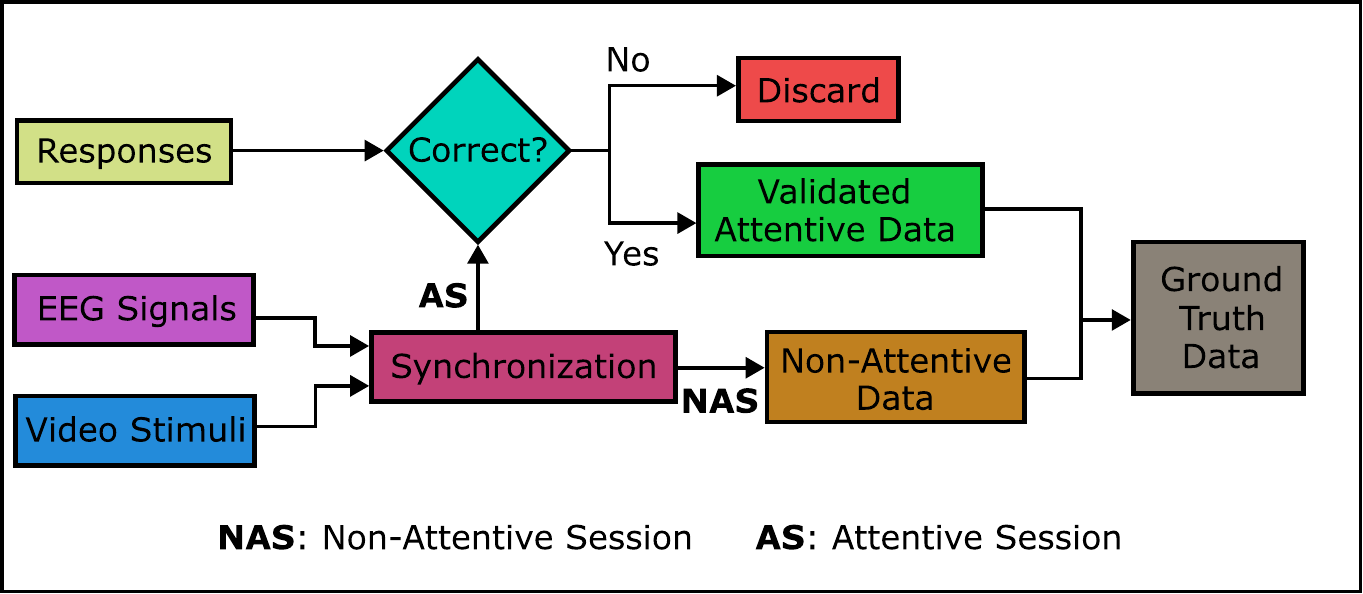}%
        \label{fig:datavalidate}}\\
    \caption{Overview of the data collection paradigm and validation procedure.}
    \label{fig:mainfig}
\end{figure*}
\subsection{Data Collection Experiment}
The data collection experiment was designed using the PsychoPy software \cite{PsychopyPeirce2019}. The design included a starting instructional video for both classes (attentive and non-attentive) that gives a brief view of the experiment done.

The participants were asked to watch a 10-minute video lecture that was split into 30-second blocks. Subsequently, a question on that exact block was shown on the screen with multiple choices, which the participants had to answer. The duration given for answering the question was set to 15 seconds, as an average human takes around 4 to 5 seconds for reading a sentence \cite{ReadingspeedKapteijns2021}. This gives the participants 10 seconds to answer which is considered ample, as humans take around 2 seconds to recognize answers \cite{RecalltimeLey1972}. The questions were set to ensure that the participant has followed the lecturer's quotes and drawings on the board. The questions require no external knowledge to be answered in order to ensure an objective evaluation of a participant's engagement. After the instructional video, a short sample video was played followed by a question-answer block to familiarize the participant with the experiment. A conditional screen was displayed with a fixation cross where the participant has to press a key to start the video lecture. In this way, each participant had to watch 20 video blocks of a lecture followed by questions inserted between them. The experimental setup captures timestamps, which mark the start and end of each video block together with participants’ responses to the questions. These records were subsequently used for data validation.

For the non-attentive data collection session, similar video sequences were presented, but rendered inaudible and blurred to prevent participants from perceiving the content. The participants were instructed to stare at the screen for 5 minutes followed by 5 minutes with eyes closed. In the eyes-open condition, participants viewed the hazy, non-informative video, which was designed as a monotonous, low-load task. Such tasks, which provide minimal external stimulation, are known to induce states of boredom or an "unengaged mind", creating a strong tendency for attention to shift from the external environment to internal thought \cite{mindwander2,mindwander1}. This method is based on extensive findings that mind-wandering is most prevalent when cognitive demands are low \cite{mindwander3}, thereby simulating a common occurrence in real-world scenarios where students face difficulties in maintaining attention during online lectures. The eyes-closed condition was designed to simulate drowsiness or dozing off during online lectures. A brief overview is illustrated in Fig. \ref{fig:datacollect}.

\subsection{Data Validation}
After successful data acquisition, the EEG data were stored with proper timestamps through some modifications of the Crimson SDK, and the responses to the questions were stored with synchronized timestamps. The correct response from a participant validates the EEG data corresponding to the video block that follows the question. All validated blocks were stored as ground truths for attentive data. For non-attentive data, the questionnaire assessment was not performed and thus required no further validation. The procedure is shown in Fig. \ref{fig:datavalidate}.

\section{Methods}
The validated data were processed through several steps, namely data pre-processing, feature extraction, and feature selection before being used as training data for classification. 
\subsection{Data Preprocessing and Preparation}
Data were pre-processed through steps that include segmentation, filtering, noise removal, and smoothing. 
\subsubsection{Segmentation}
EEG data of each participant of a total length of \( T \) samples were partitioned to get data segments of optimal length \( L= 1750 \) samples (details in section III.D) . Although the average sampling frequency of 250 Hz was retained, samples were used instead of time intervals to avoid errors caused by minute temporal delays in packet transmission, which complicate timestamp-based calculations. Through validation, a total of \( K \) blocks were selected from the initially acquired \( n \) blocks for each participant. This number corresponds to the average number of questions (around 20) set for a video lecture per session. A sliding window was used during segmentation to reduce the edge effects and improve statistical reliability \cite{OverlapSaideepthi2023}. The sliding window is also referred to in this article as the overlap ratio \( r=0.7 \) (details in section III.D). The total number of segments found for each selected block is:
\[
n_k = \left\lfloor \frac{T_k - L}{L(1 - r)} \right\rfloor + 1
\]
Thus, the total number of segments found is:
\[
n = \sum_{k=1}^{K} n_k = \sum_{k=1}^{K} \left( \left\lfloor \frac{T_k - L}{L(1 - r)} \right\rfloor + 1 \right)
\]

The total number of samples is $x = n \times L$.
During segmentation, two columns were added for labeling and assigning subject-wise user\_ids. These were used later in classification.

To ensure subject-wise label balance, samples for both classes were extracted separately for each subject. Let \( \mathcal{D}_u^0 \) and \( \mathcal{D}_u^1 \) denote the sets of samples with labels 0 and 1 respectively (0 meaning attentive and 1 meaning non-attentive), for a given user \( u \). An equal number of samples was retained by trimming the larger class to match the size of the smaller one. Number of samples to be retained, \( n_u = \min(|\mathcal{D}_u^0|, |\mathcal{D}_u^1|) \), and the last \( n_u \) samples from each label set were selected, which gave \( \mathcal{D}_{u,\text{tail}}^0(n_u) \) and \( \mathcal{D}_{u,\text{tail}}^1(n_u) \). The balanced dataset for user \( u \) was then
\[
\mathcal{B}_u = \mathcal{D}_{u,\text{tail}}^0(n_u) \cup \mathcal{D}_{u,\text{tail}}^1(n_u),\] and the final dataset was\[  \mathcal{B} = \bigcup_{u \in \mathcal{U}} \mathcal{B}_u\]
Given that the dataset retained a substantial number of segments, class imbalance was addressed by trimming the larger class rather than oversampling the smaller one.

\begin{figure*}[!t]
    \centering
    \subfloat[Filtering without Trimming]{%
        \includegraphics[width=0.45\textwidth]{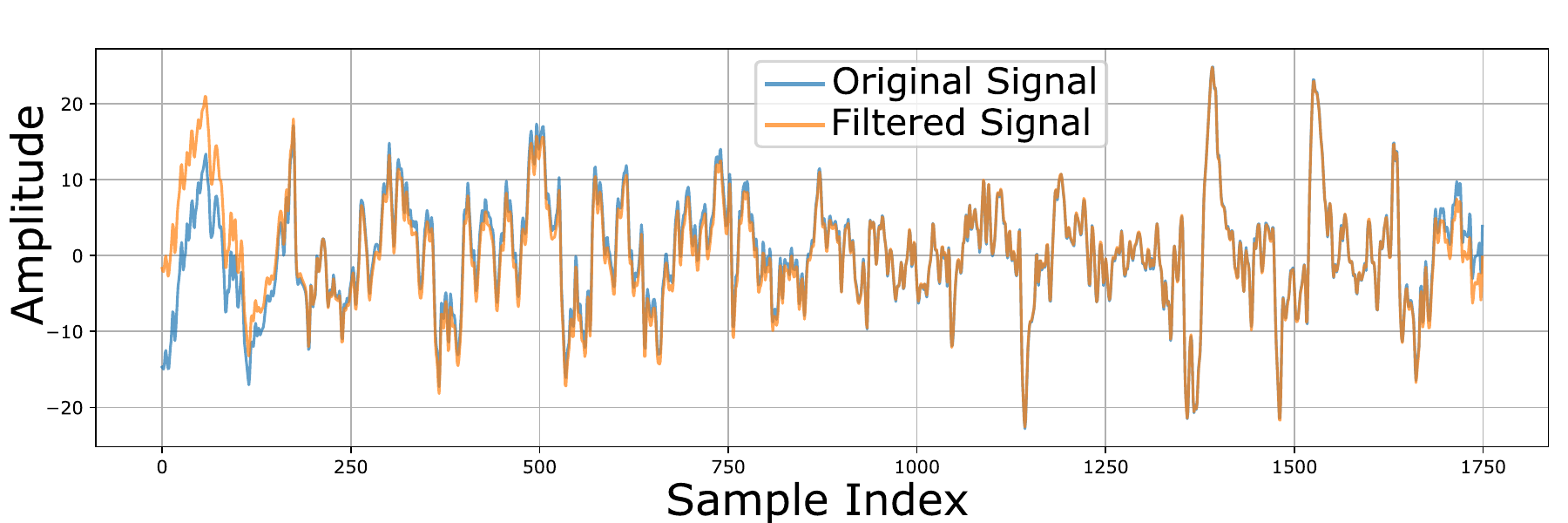}%
        \label{fig:filtnotrim}}
    \subfloat[Filtering with Trimming]{%
        \includegraphics[width=0.45\textwidth]{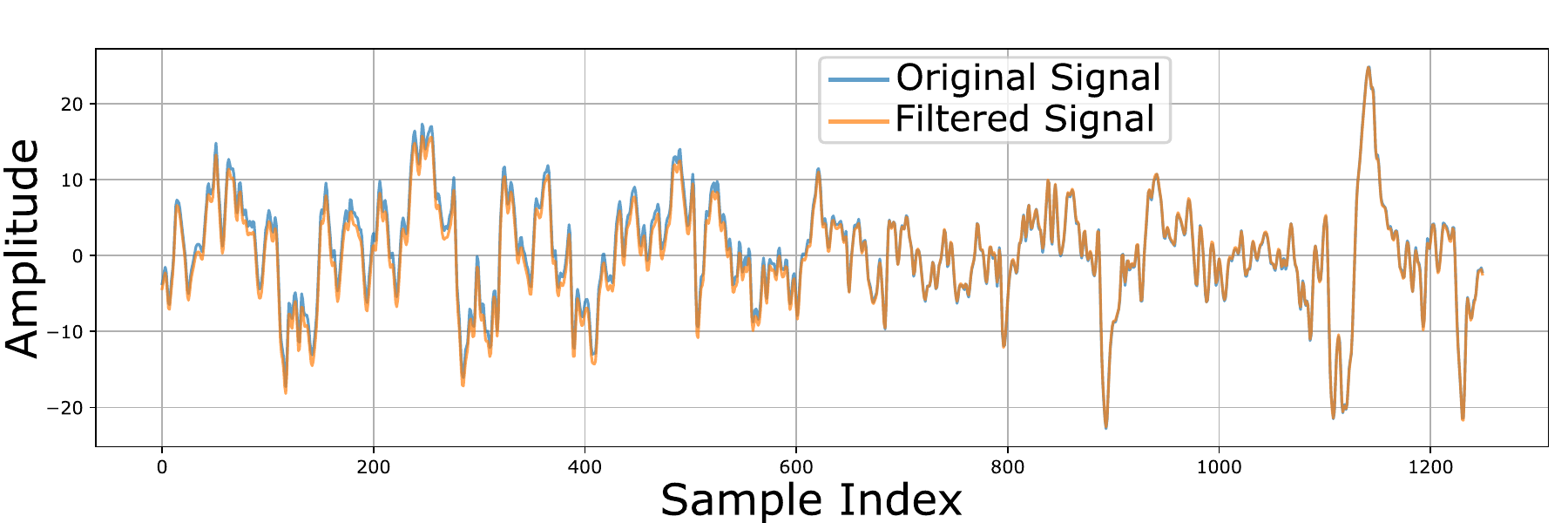}%
        \label{fig:filttrim}}\\

    \subfloat[Data Segmentation with sliding window]{%
        \includegraphics[width=0.9\textwidth]{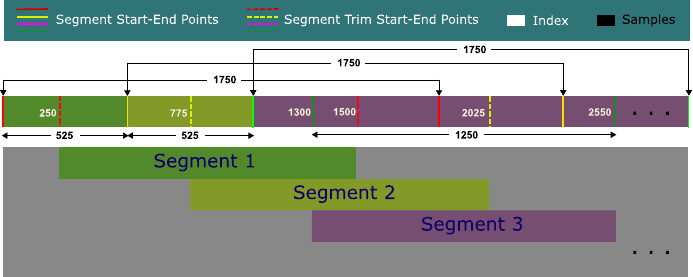}%
        \label{fig:segmentation}}

    \caption{Subfigures (a) and (b) illustrate the effect of trimming on filtered EEG signals, showing signals without trimming in (a) and with trimming in (b). Subfigure (c) presents the data segmentation pipeline using overlapping windows, highlighting how trimming affects the resulting segments.}
    \label{fig:mainfig}
\end{figure*}

\subsubsection{Filtering}
The EEG signals were filtered using a bidirectional Butterworth bandpass with a cutoff frequency of 0.5–64 Hz, which is a standard range in this research field, and an optimal order of 3 found through graphical comparisons. The problem arises due to edge effects caused by filter initialization and group delay as seen in Fig. \ref{fig:filtnotrim}. To overcome this, 250 samples were trimmed from both ends of each segment to get clean signal with minimal phase distortion after filtering. The procedure is illustrated in Fig. \ref{fig:filttrim}. This method added a constraint to the selection of segment length for experimentation which is $L>500$ samples. This means that the new segments found after trimming had a length of (\( L-500 \)) samples, which gave rise to another constraint of setting the ratio \( r \). The constraint can be understood as  $r \times L>500$. The logical explanation is understandable from the overall segmentation and trimming procedure, depicted in Fig. \ref{fig:segmentation}, where the sequential overlapping and trimming of the segments are shown along with the obtained segments. Finally, the signal was passed through a 50 Hz notch filter to remove power line interference. 
\subsubsection{Artifact Removal}
Non-invasive EEG signals are often contaminated by artifacts, including eye blinks. Two steps were performed for noise removal. First, the filtered segments were checked for high-amplitude values greater than a threshold of 150 $\mu\text{V}$, as meaningful EEG data lie below this value \cite{microvolt150}. If any segment contained at least one high amplitude value, then that segment was dropped from the dataset. Correlation and interpolation are commonly used for high-amplitude portions \cite{Correlationuse,Artifacts}, but the segment rejection approach was used to exclude artificial continuations, given the abundance of collected data. Following this, eye blinks were removed through Ensemble Empirical Mode Decomposition (EEMD) \cite{EEMD}. Peak indices were extracted from the filtered EEG signal using the find\_peaks function from the SciPy library \cite{Scipy}, which finds all the local maxima. Then three conditions were imposed to take the region around the peak. If a peak index fell within the starting 40 or ending 80 values of the segment, then the starting 120 or ending 120 values were taken for EEMD. However, if the peak index fell within the rest of the region then the 40 values before and 80 values after the peak index were taken. These indices were acquired through rigorous experimentation to get optimal artifact removal. The signal found after performing EEMD was then smoothed using a uniform filter to eliminate residual high-frequency noise.
\subsection{Feature Extraction}
Following pre-processing, feature extraction was applied to the segments. 
A comprehensive set of 458 features was extracted from the EEG signals to capture time-domain, frequency-domain, wavelet-domain, and statistical characteristics. Time-domain features included commonly used descriptors such as  Hjorth parameters (e.g., complexity, activity), median (md), normalized second difference (n2d) etc. Frequency-domain features encompassed spectral moments (e.g., centroid, spread), entropy-based measures, and spectral shape descriptors (e.g., crest, flatness, roll-off). Additionally, wavelet packet transform (WPT) was used to extract sub-band energy and entropy features, which effectively captures localized time–frequency dynamics. These features were selected based on their proven effectiveness in attention evaluation through EEG and biosignal analysis. Eight additional features from the Crimson SDK (six frequency band powers along with attention and meditation scores) were appended to the initial 458 features, for a total of 466. These standard, open-source features were included to explore their potential contribution. Following feature extraction, a data cleaning step was performed. Any features containing NaN or Inf values (arise from divisions by zero or logarithmic transformations of zero-valued data) were replaced with zero. Zero was chosen as a neutral, non-informative value that preserves data dimensionality and ensures numerical stability for model training.

\subsection{Feature Selection}
The next step after feature extraction was selection of the most relevant features for training the data. This was necessary to ensure that the final model was trained on the best set of features that contribute to the classification performance. The selection was done in two steps: Pearson correlation filtering (PCF), followed by Recursive Feature Elimination (RFE) using Support Vector Machine (SVM). 
\subsubsection{Pearson Correlation Filtering}
PCF was used to reduce high dimensionality and optimize computational efficiency, which works by selecting one from each highly correlated pair when the coefficient of correlation exceeds a threshold $p$. The value of $p=0.8$ (details in section III.D) was used to remove the highly correlated features, leaving 108 in the set for further analysis. The Pearson correlation coefficient is defined as:
\[
r_{xy} = \frac{\sum_{i=1}^{n} (x_i - \bar{x})(y_i - \bar{y})}
{\sqrt{\sum_{i=1}^{n} (x_i - \bar{x})^2} \sqrt{\sum_{i=1}^{n} (y_i - \bar{y})^2}}
\]
where $x$ and $y$ are the features, $\bar{x}$,$\bar{y}$ are their means and $n$ represents the number of paired observations.
\subsubsection{Recursive Feature Elimination with SVM}
Subsequently, RFE with SVM was applied to find the meaningful set of features required for optimal model training. It is a supervised feature selection method that removes the least important features recursively based on the weights assigned by a linear SVM model. At each iteration, the model was trained on the current feature set, and features with the smallest absolute weights (least contribution to the decision function) were eliminated. This process was continued until the desired number of features, $d= 50$ (details in section III.D), was retained.

First, RFE was applied for each subject (identified by user\_id) using a linear SVM. The regularization parameter $C_1$ for this SVMwas iterated over a range (0.01 to 100), to account for varied model complexity and feature importance. For each $C_1$ value, a set of 50 top-ranked features was obtained. These sets were then evaluated through another SVM with a radial basis function (RBF) kernel. Classification performance of this secondary model served as the basis for selecting the optimal feature set. This subject-focused process was designed so that the selected features were not only ranked high based on linear separability but also demonstrated better generalization of the model. Finally, a set of 9 common features was obtained by finding the most frequently selected features across all subjects. These included the relative powers of alpha and delta band, ratio of low beta to summation of alpha and theta band, average power of low beta band, and median of high beta band. These features showed strong correlations with those reported in the literature as contributing to sustained attention \cite{frontalbetaalsoKaushik2022,frontalregion1Sahu2024,frontalZhang2021,betabandpowerKo2017}. These consensus-based features were used for final model training. The selected features for final training along with their meanings are given in Table \ref{tab:selectedfeatures}.

\begin{table}[h]
  \caption{Selected Features for Training}
  \centering
  \begin{tabular}{|c|c|}
    \hline
    Feature & Meaning \\
    \hline
    RP\_A & Relative power of alpha band \\
    RP\_D & Relative power of delta band  \\
    attention & Attention Score from Crimson SDK \\
    avg\_pow\_B1 & Average power of low beta band\\
    en\_b\_at & Ratio of low beta to summation of alpha and theta band\\
    hc\_D & Hjorth complexity of delta band\\
    md\_B2 & Median of high beta band\\
    meditation & Meditation Score from Crimson SDK\\
    n2d\_G & Normalized second difference of gamma band\\
    \hline
  \end{tabular}

  \label{tab:selectedfeatures}
\end{table}

\subsection{Hyperparameter Tuning}
All parameters defined from pre-processing to feature selection were systematically varied to evaluate their impact on classification performance through rigorous assessments. For the segment length L, values ranging from 5 to 8, paired with overlap ratios ranging from 0.3 to 0.7, were used for analysis. The pair of $(L,r)=(1750,0.7)$ was found to be optimal through comparisons of model performance across all tested values. A similar approach was followed for the Pearson correlation coefficient threshold $P$ (0.6 to 0.95 in increments of 0.05) and the optimal number of features $d$ (30 to 80 in increments of 10). The best trade-off between model performance and generalization was found for $P=0.8$ and $d=50$.  
For RFE, the linear SVM was trained on $C_1 \in \{0.01,\ 0.1,\ 1,\ 10,\ 100\}$,  and the secondary SVM for identifying the optimal feature set was trained using a nested grid search over two parameters: $C_2 \in \{0.01,\ 0.1,\ 1,\ 10,\ 100\}$ and $\gamma \in \{0.001,\ 0.01,\ 0.1,\ 0.5, \ 1,\ 10\}$. The combination yielding the highest performance determined the best feature set corresponding to the optimal $C_1$. Final classification was further tuned using a separate grid search on the same range of $C_2$ and $\gamma$, applied to the common features obtained across all users. The optimal pair $(C_2,\gamma)=(0.01,0.5)$ was found to yield the best classification performance.

\subsection{Classification}
SVMs are widely adopted for EEG classification tasks due to their robustness in high-dimensional spaces, effectiveness with small sample sizes, and ability to model non-linear decision boundaries through kernel functions. Therefore, an SVM with an RBF kernel was selected to capture non-linear relationships in the feature space. The SVM finds a hyperplane that maximizes the margin between classes while allowing some misclassifications, which are controlled by the regularization parameter ${C_2}$. The kernel coefficient $\gamma$ determines the influence of individual training samples on the decision boundary, thereby shaping how tightly the model fits local patterns in the feature space.
The training pipeline included fitting the model on a nested grid search of $C_2$ and $\gamma$. The evaluation was performed through leave-one-subject-out (LOSO) cross-validation. LOSO is an appropriate choice to effectively assess the model’s generalizability to unseen subjects, thereby closely reflecting real-world application scenarios.


    

After a rigorous grid search, the pairs of $C_2$ and $\gamma$ yielding the best classification performance for the subjects were recorded. From these pairs, the most frequent combination was selected to fit the final model, which was then used for real-time evaluation.
\subsection{Evaluation Metrics}

    
    
    

The performance of the classifier was evaluated using standard metrics, including accuracy, precision, recall, and F1-score. Since the dataset was balanced per class during segmentation, these metrics were sufficient to estimate the model performance.
\begin{figure*}[h]
    \centering
    \includegraphics[width=0.9\linewidth]{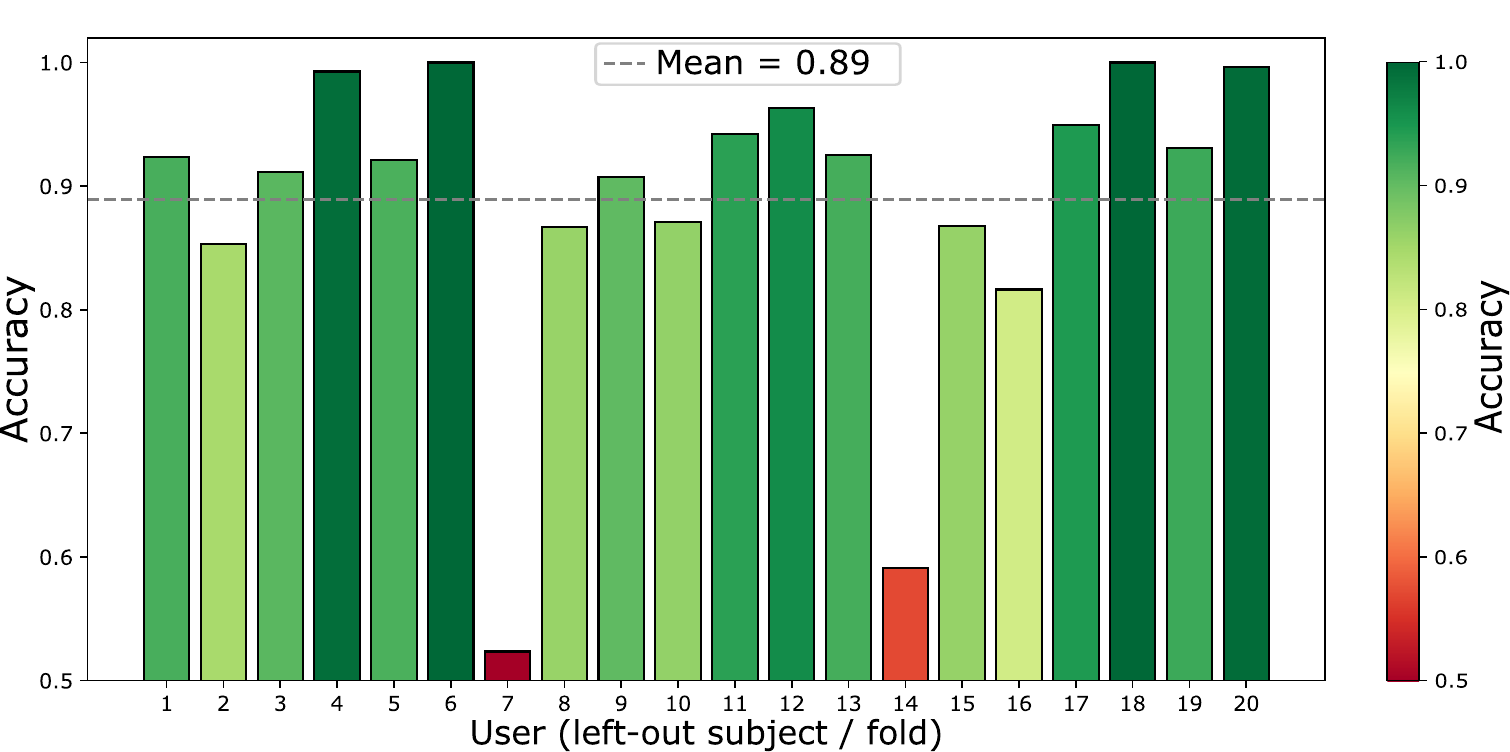}
    \caption{Accuracy of the model evaluated using Leave-One-Subject-Out (LOSO) cross-validation, where each iteration treats one subject as the test fold and the remaining subjects as the training set.}
    \label{fig:losostats}
\end{figure*}
\section{Experimental Results}
This section presents the performance evaluation of the trained model and the ablation studies that led to its optimized version.
\subsection{Cross-Validation Results}

The model performance was analyzed using the LOSO cross-validation procedure (discussed in section III.E). The optimized model's performance on each fold (fold implying the data of one subject as test data while taking all the others as train data) is shown in Fig. \ref{fig:losostats}. The optimized model achieved an average test accuracy of 88.77\%. Additionally, it attained a precision of 90.49\%, recall of 88.77\%, and F1-score of 89.03\%, which indicates strong predictive capability with a good balance between correctly identifying positive cases and minimizing false positives.

\subsection{Model Ablation Studies}
Initially, 458 features excluding the 8 features provided by the Crimson SDK, were used to train the SVM, which gave an accuracy of 76.44\%. Subsequently, the eight SDK features were incorporated, and the model was retrained, achieving the current best performance. Among these additional features, the attention and meditation scores were included in the final feature set. To assess whether the model’s performance is overly reliant on these two scores, it was compared against the baseline performance obtained using only the Crimson SDK’s scores as the feature set. The Crimson SDK was selected as the baseline, as it provides the scores associated with cognitive load. To evaluate the predictive accuracy of these scores for the EEG signals, two methods were employed. The first method included training the model on the two features and evaluating the metrics, which gave an accuracy of 84.13\%. The second method involved the ratio of attention to meditation score $r_{am}$ as the condition to classify the data. For all the n segments found, the predictions were taken to be attentive if $r_{am}>1$ and non-attentive otherwise. The accuracy was found to be 82.27\%. This shows that the proposed methodology outperforms the current interface of the FocusCalm headband modified for attention classification in online learning environments (details discussed in section V). A brief summary is provided in Table \ref{tab:ablation} for a clear overview. 
\begin{table}[h]
  \caption{Model Ablation Performance}
  \centering
  \begin{tabular}{|c|c|}
    \hline
    Method & Performance \\
    \hline
    458 features excluding attention and meditation scores& 76.44\% \\
     \hline
    Ratio of attention and Meditation scores & 82.27\%  \\
     \hline
    Attention and meditation scores as final features & 84.13\% \\
     \hline
    466 features including attention and meditation scores& 88.77\%\\
    \hline
  \end{tabular}
  \label{tab:ablation}
\end{table}

\section{Discussion}
The overall research methodology differs from baseline methods (given the limited prior research addressing this objective within this specific environment). Therefore, the discussion focuses on the unique aspects of the approach, the rationale for their selection, their implementation, and the specific methods they outperform. Finally, this section addresses the limitations of the current research and provides a logical justification for the methods selected.

The proposed EEG-based attention classification achieves 6.5\% better accuracy than the Crimson SDK, which suggests that the combination of wavelet packet decomposition along with entropy features are effective in capturing attention state dynamics. Although baseline methods achieve satisfactory results, the data collection strategies of other baseline studies such as Rehman et al. and Huang et al. are based on instructions and there was room for data validation strategies which are implemented in this research. The steps followed in this research ensure correct predictions which pertain to ground truths and this is crucial for real-time implementation. 

Initially, device selection was emphasized, as most previous research relied on non-consumer-grade devices. For this study, the FocusCalm headband was chosen for its portability, wearability, and ease of hands-on use, despite the potential for experiments with the Muse headband, which offers greater spatial coverage in real-time applications. Given our objective of developing a state-of-the-art real-time binary classification system, the FocusCalm device has proven sufficient for our purposes. The data collection protocol was designed to reflect the environmental setting of classrooms, with video stimuli consisting of lecture content prepared for students. The data validation process was established under the principle that only ground-truth labels are required. Instances can occur where a participant experiences cognitive load yet fails to answer a question. However, this did not affect labeling, as the protocol ensures that all selected data correspond to verified ground truths. 

The methods were selected through extensive trials. For instance, both causal forward filtering and zero-phase second-order section (SOS) filtering were tested. While zero-phase SOS filtering was more effective at removing phase distortion, its non-causal nature performed poorly at segment edges. To address this, those edge portions were removed, ensuring both the absence of distortion and the independence from future data. Furthermore, to prevent data loss, the segmentation process used an overlapping ratio. For classification, SVM was initially selected due to its proven effectiveness in EEG signal analysis. Conceptually, SVM is viewed as a mechanism that iteratively identifies the most discriminative features in a high-dimensional space, analogous to how traditional EEG analyses iteratively isolate regions or frequency bands of interest to distinguish cognitive states. An initial assessment using various machine learning models was conducted, and the SVM achieved the highest performance, which is consistent with observations reported in the literature.

After validating the data using LOSO, the optimal $(C_2,\gamma)$ values were determined in two ways. The first method identified the most common $(C_2,\gamma)$ pair that achieved the best metrics across all subjects. The second method selected the $(C_2,\gamma)$ pair that yielded the highest metrics, regardless of its frequency among subjects. The mode pair, $(C_2,\gamma) = (0.01, 0.5)$, produced an average accuracy of 88.77\%, whereas $(C_2,\gamma) = (0.01, 0.1)$ achieved a slightly higher average accuracy of 89.63\%. Despite the higher accuracy, the former pair was adopted for real-time testing due to its more frequent favorable performance, which makes the model generalized and objective. The LOSO validation results given in Fig. \ref{fig:losostats} show that while the model performed well on most subjects, it performed poorly on two. This indicates the model's robustness could be improved with additional data.
\begin{figure*}[!t]
    \centering
    \subfloat[Graphical user interface (GUI) for real-time EEG data visualization and classification.]{%
        \includegraphics[trim=0 0 0 5,clip, width=0.5\linewidth]{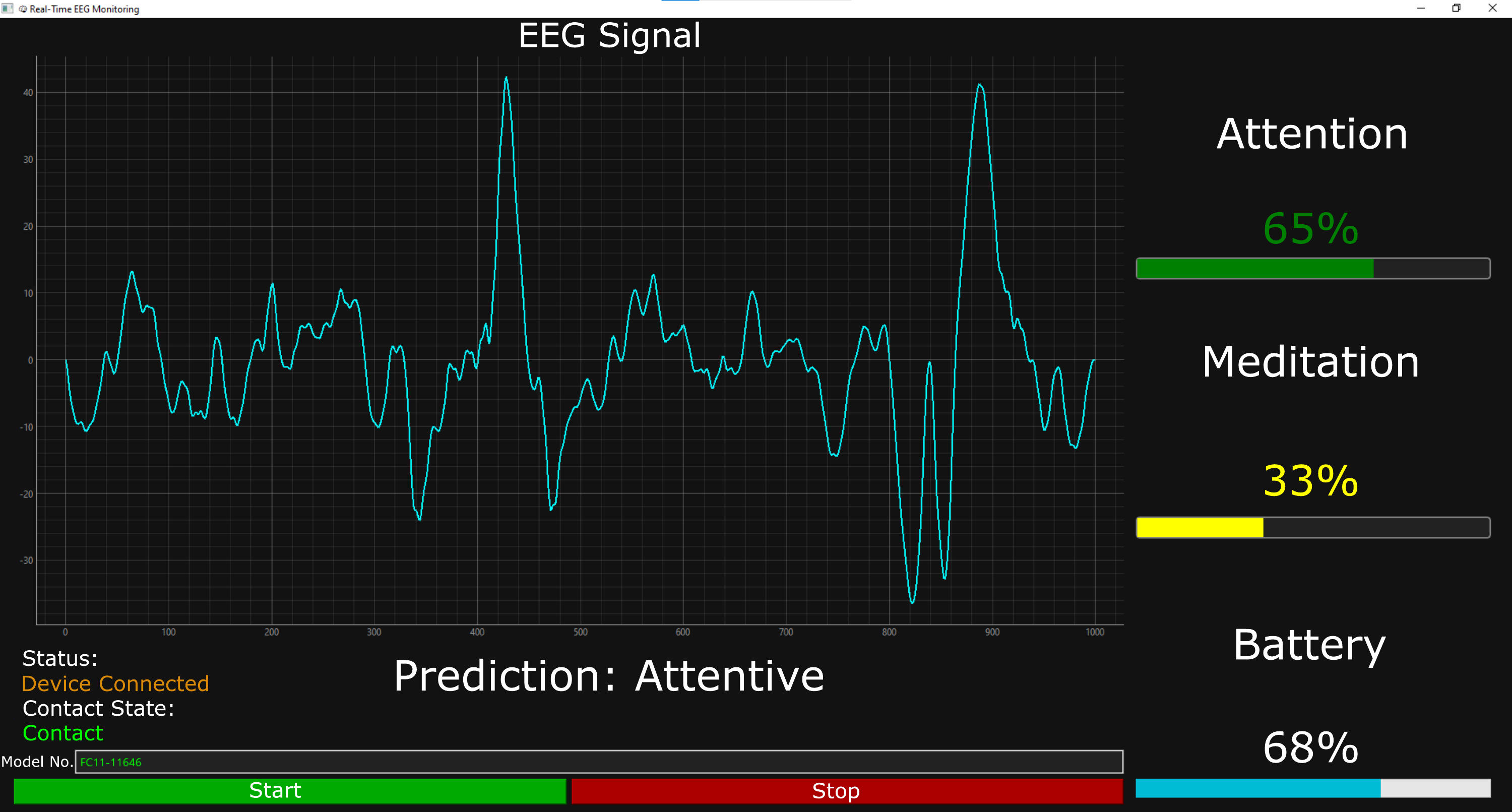}
        \label{fig:GUI} 
        }
    \hspace{0.02\textwidth} 
    \subfloat[Overview of the online learning pipeline, showing EEG signal acquisition, real-time classification, and neurofeedback to regain attention.]{%
    \includegraphics[width=0.35\linewidth]{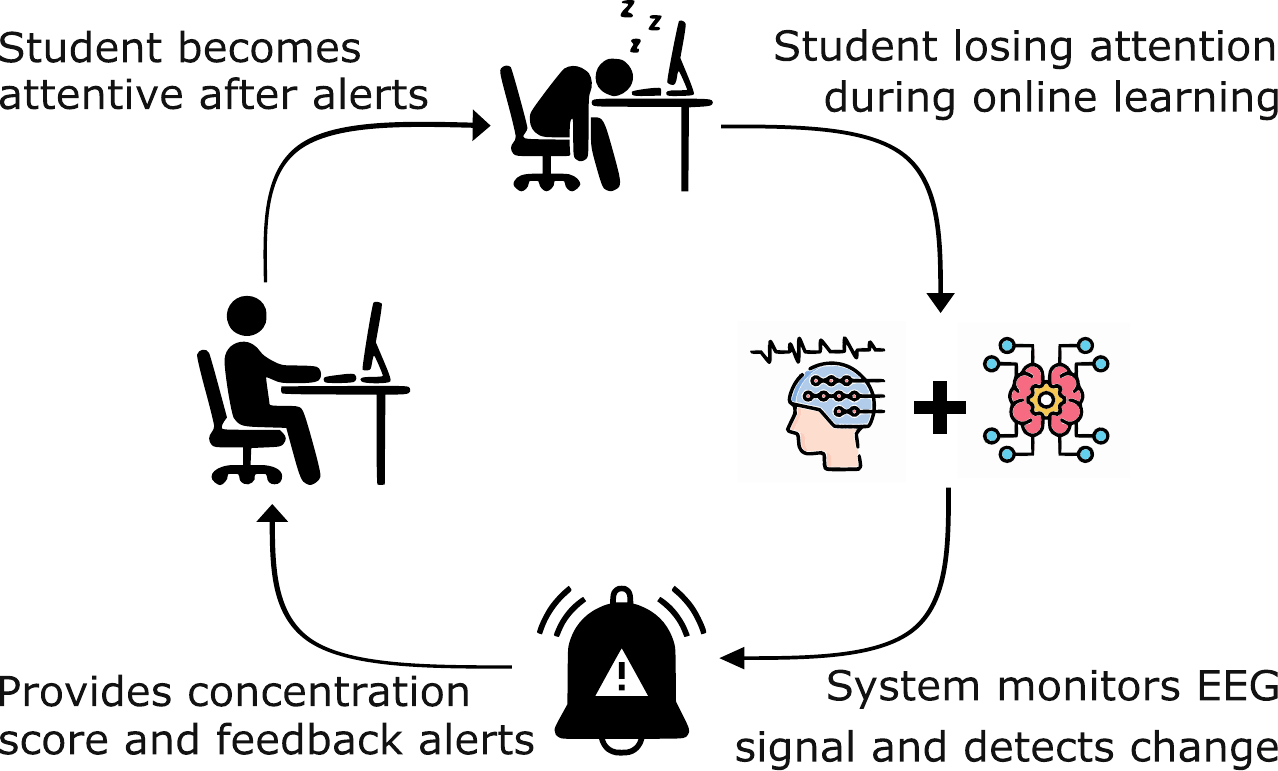}
    \label{fig:pipeline}       
        }\\

    \subfloat[Graphical visualization of cognitive states in two phases of conducted pilot study.]{%
        \includegraphics[width=0.9\textwidth]{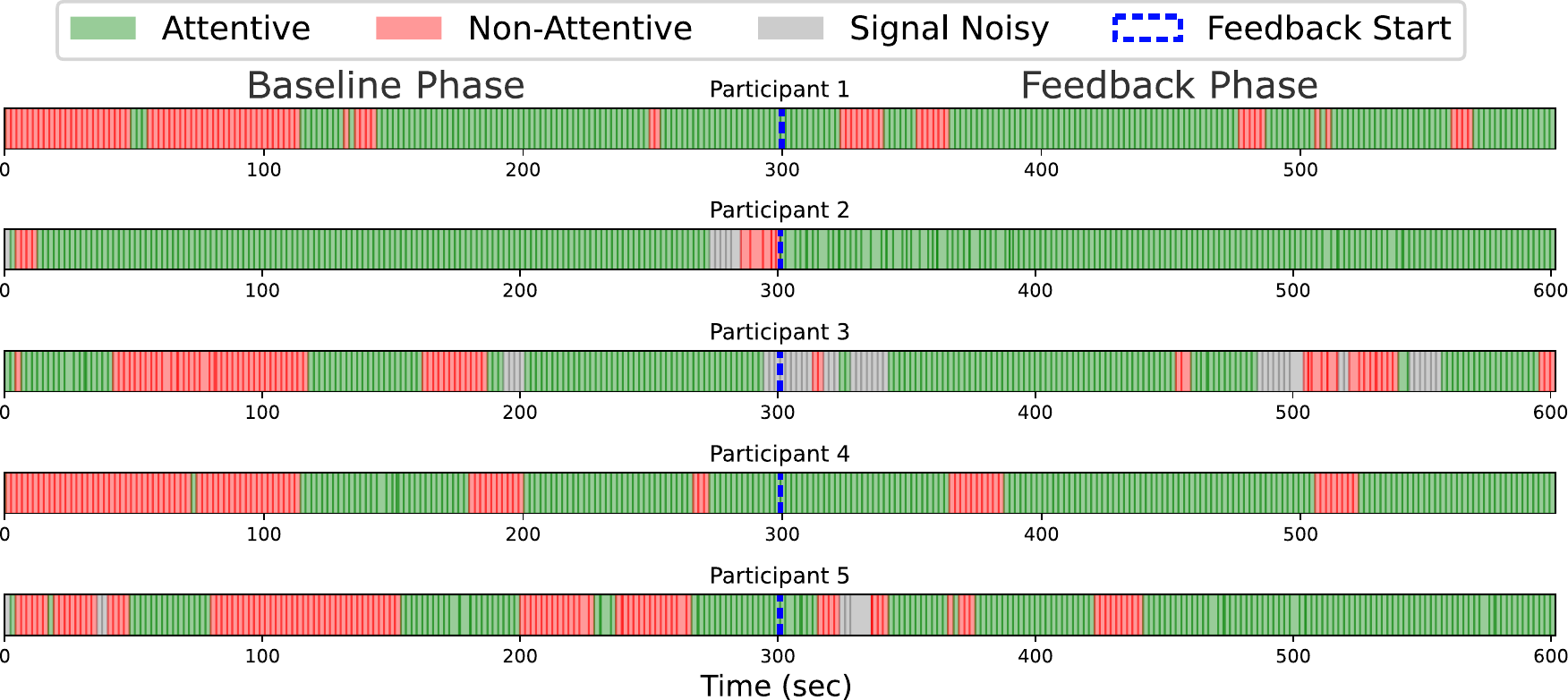}%
        \label{fig:realtimeplot}}

    \caption{Subfigures (a) and (b) represent the graphical user interface and the real-time working pipeline. Subfigure (c) denotes the changes between attentive and non-attentive states during the real-time system test on five participants.}
    \label{fig:mainfig2}
\end{figure*}
\section{Real-time Setup}
\subsection{Data-handling Procedure}
The real-time system was implemented in Python using a circular buffer for continuous EEG data processing. Fixed-length segments of 1750 samples were extracted with a step size of 525 samples (70\% overlap), resulting in a new classification output approximately every 2.1 seconds. Each segment was processed using noise removal, feature extraction, and classification by the trained model.
\vspace{-2mm}
\subsection{Graphical User Interface Development}
The graphical user interface (GUI), developed with PyQt6, displays real-time EEG signals, SDK metrics, the algorithm's prediction, and headband status. The GUI is shown in Fig. \ref{fig:GUI}. The system provides real-time feedback: an audio-visual alert is triggered after five consecutive non-attention segments. Similar warnings prompt the user to adjust the headband if contact is lost or if significant artifacts are detected. At the end of a session, a summary screen displays detailed attention statistics. This feedback loop is visualized in Fig. \ref{fig:pipeline}, which aims to promote cognitive engagement. 

\subsection{Pilot Evaluation}
To preliminarily assess the effectiveness of the implemented feedback mechanism, a pilot study was conducted involving five participants. Each participant watched a 10-minute educational video. The video was split in two consecutive phases: a 5-minute baseline phase without feedback and a 5-minute feedback phase. During the baseline phase, the system continuously monitored cognitive states without alerts. For the feedback phase, the same monitoring was done and the system issued visual and audio alerts whenever a participant remained non-attentive as per the model for approximately 8.4 consecutive seconds (4 overlapped segments).

A paired t-test was conducted on the mean durations of non-attentive states in the two phases. At first, non-attentive segments with less than 8 seconds duration were discarded as the effect of feedback mechanism was being evaluated. The paired t-test yielded a t-statistic of 5.73, and a p-value of 0.007. Moreover, the mean duration of non-attention periods for the participants was found to be 37.50 seconds in the baseline phases and 14.97 seconds in the feedback phases. Both analyses indicated a significant improvement in concentration during the feedback phase, suggesting that the real-time alerts effectively supported participants in regaining and sustaining attention. Fig. \ref{fig:realtimeplot} gives a graphical overview for better visualization, portraying the timeline of sustained non-attention in the baseline phase and the feedback phase. 

        
\section{Conclusion}
As traditional methods like questionnaire and observations require manual interventions, they are not suitable for online learning. For this reason, an automated real-time attention classification with neurofeedback mechanism has been developed in this article. The methodology, based on data from the FocusCalm headband, employed advanced preprocessing, WPT-based feature extraction, and an SVM classifier. The proposed algorithm achieved an average accuracy of 88.77\% on leave-one-subject-out validation. This result demonstrates the robustness of the model on unseen subjects and proves the feasibility of using consumer-grade BCI for real-time attention monitoring.

Despite these promising results, some limitations are acknowledged. The model's performance was validated on a limited participant pool. Furthermore, the optimal feature set includes proprietary scores from the Crimson SDK, which may limit the model's transparency and adaptability.

Future work will directly address these limitations. We will first focus on expanding the dataset to include a more diverse participant pool. Better and more robust methods will be developed to ensure better performance of the model without requiring proprietary Crimson SDK scores. The application will be analyzed through implementation in multiple use cases, followed by systematic result collection. The use of multi-channel, consumer-grade and ready-to-use EEG devices will be adopted for better spatial coverage to capture and classify visual, auditory, and task distractions alongside the current attention and non-attention states. Another possible implication is integrating this research with multi-modal inputs such as facial expressions and behavioral logs, to evaluate the key dimensions of a class: learner engagement and content quality. This research thus paves the way for practical and scalable real-time BCI applications.

\section*{Acknowledgments}
We gratefully acknowledge the Advanced Intelligent Multidisciplinary Systems Lab, under the Institute of Research, Innovation, Incubation \& Commercialization at United International University, for their support and resources, Institute of Advanced Research for funding this project (Code No. UIU-IAR-02-2024-SE-39) and Onnorokom EdTech Limited for testing our project in their learning environment.




\vspace{11pt}


 




\vfill

\end{document}